# Probing the Valley-Selective Tunneling Density of States in Monolayer-$MoS_2$ based Resonant Tunneling Devices

**Abir Mukherjee[1], Kajal Sharma[1], Ajit K Katiyar[2,3], Saranya Das[4], Samit K Ray[4] and Samaresh Das[1*]**

[1] Centre for Applied Research in Electronics, Indian Institute of Technology Delhi, Delhi, 110016, India

[2] School of Electrical and Electronic Engineering, Yonsei University, Seoul 03722, Republic of Korea

[3] Dept. of Physics, School of Advanced Engineering, UPES Dehradun, Uttarakhand 248007, India

[4] Dept. of Physics, Indian Institute of Technology Kharagpur, Kharagpur 721302, India

*** Corresponding Author: samareshdas@care.iitd.ac.in**

## Abstract

The present work experimentally demonstrates the fabrication of CVD grown monolayer $MoS_2$ ultra-thin quantum well based double barrier resonant tunneling device (RTD) architecture well compatible with conventional CMOS fabrication technology. The strongly quantized electronic states from multiple valleys in the momentum space in such ultra 2D-sheet along the c-axis sandwiched in between $Al_2O_3$ tunneling barriers exhibit multiple resonant tunneling peaks thereby enhancing the FWHM of the NDR region as derived from experimental I-V characteristics as well as theoretical joint invision through Density Functional Theory (DFT) and Non-Equilibrium Green's function (NEGF) visualized via Tunneling Density of States (TDOS). Understanding extended to S-vacancies not only change the bandgap, as evaluated through nanoscale Cathodoluminescence (CL) spectroscopy, but also alters the effective mass hence the mobility as investigated here within the high symmetry path in the k-space. Electrical performances of fabricated RTD, starting from cryogenic to room temperatures, show a significant milestone via exhibiting huge PVR values of 178 at 4K and 24 at RT with more possible improvement in the field of room temperature quantum technology. Momentum conserved and non-conserved tunneling from highly n-doped Si through multiple valleys of 1L-$MoS_2$ provides a tremendous opportunity in gate-induced manipulation in Spin-Valley Qubit technology operational at deep cryogenic temperatures (mK).

# I. Introduction

The advent of quantum computation and quantum information processing in practice is fundamentally based upon the development of qubits, where quantum superposition principle leads to coherent oscillation between two qubit-states within their characteristic coherence time[1–5]. In the construction of spin/charge qubit, fundamental inherent understanding is all around resonant tunneling and its strength characterized by the peak to valley ratio (PVR)[5–10]. In the field of resonant tunneling devices, III-V semiconductors (e.g. GaAs, InP, GaN etc.) show the dominant hold via demonstrating: 1. Lower effective mass i.e. higher tunneling probability, 2. Sharper and stronger negative differential resistance (NDR), 3. Higher peak current density with atomically abrupt, lattice-matched interfaces such as GaAs / AlGaAs, InGaAs/AlAs etc[11–14]. The emergence of two-dimensional (2D) materials, coupled with advancing fabrication techniques for stacking 2D materials, has opened numerous pathways to explore the electronic and photonic properties, and device applications[9,10,15–19]. Evolving techniques for the layer-by-layer transfer of 2D materials allow for great flexibility in device structure and have led to the study of many interesting phenomena utilizing different van der Waals architectures (vdW), such as quantum Hall effect and moiré-bands in high mobility graphene on h-BN substrates, quantum Hall effect in TMDs encapsulated with h-BN and resonant tunneling in double monolayer or double bilayer graphene separated by h-BN[20–23], the relative rotational alignment of different layers is most often not controlled. Because resonant tunneling requires a precise overlap of states in momentum space, and desirably a strong interlayer coupling, it serves as a powerful tool to probe the quantum fingerprints of vertical transport in vdW heterostructures. Furthermore, the gate-tunable NDR of the interlayer current−voltage characteristics enable the implementation of novel interlayer tunneling field-effect transistors (ITFETs) with applications for both Boolean and non-Boolean logic[24–26]. Molybdenum disulfide ($MoS_2$) is a prototypical transition metal dichalcogenide which has a layered structure wherein molybdenum atoms are sandwiched between the layers of sulfur atoms in a hexagonal arrangement. Atomic $MoS_2$ sheet is a semiconductor with large bandgap. Like graphene, single layer $MoS_2$ possesses unique properties such as high transmittance and two-dimensional (2D) flexible geometry and has become an emerging material in the field of nanoelectronics. 2D-RTD still suffers over III-V RTDs by means of room temperature application and conventional established CMOS fabrication techniques[9,20,27]. For comprehensive understanding of the transport behavior of such devices, it is essential to develop quantum mechanical models which can incorporate the effects of discontinuities in the energy continuum and consequently the variation of the relevant effective mass components. Such a model can be obtained by developing a Schrodinger-Poisson simultaneous solver which can help in visualization of the quantum transport of the charges. In this context, the nonequilibrium Green's function (NEGF) formalism has appeared as one of the most potential mechanisms that can solve the quantum transport equations of electron in the active device, considering its interaction with various types of reservoirs such as light source, phonons, electrical & magnetic reservoirs etc. Several reports are available where such techniques have already been used to study the transport behavior of different nano-scale MOSFET architectures such as double gate MOSFET, Fin-FET, multiple gate NWFET, gate-all-around (GAA) NWFET, nanotube FET, and molecular switch junctions[28–30]. However, the transport effective masses of the active device and reservoir regions are not identical due to difference in their doping concentration as well as involvement of electrons in transmission from different band minimas from E-k diagrams. Hence it modifies their interactions, leading to the change of transport behavior and electrical characteristics of the device. This work demonstrates the fabrication of 2D-RTD correlating with conventional TMDC growth technique and CMOS fabrication, well-comparable NDR characteristics from room temperature to cryogenic temperatures (≥ 4K), followed by the electrical-characterization and illustration of non-equilibrium transport physics involved.

## II. Material Characterization and Device Fabrication

In this article, to fabricate a double barrier resonant tunneling structure, monolayer-$MoS_2$ was chosen as the tunneling medium. Wafer scale large monolayer $MoS_2$ was grown via a 3-zone chemical vapor deposition (CVD) system under low vacuum condition ~ 4.5 x $10^{-3}$ mbar. To prevent growth-induced damage or defect formation in the tunneling barriers (TBs), $MoS_2$ film was grown on $SiO_2$/Si substrates rather than directly on the barrier layers, enabling subsequent transfer onto the desired substrates. Here, wet transfer methods have been carried out. A mixed etchant solution of deionized (DI) water and buffered hydrofluoric acid (B-HF) in a 20:1 ratio was prepared to selectively etch the $SiO_2$ layer and release the $MoS_2$ film from the substrate (Figure 1(d)). Direct exposure of ultrathin $MoS_2$ to the etchant can lead to severe damage or even complete dissolution due to its atomic-scale thickness.

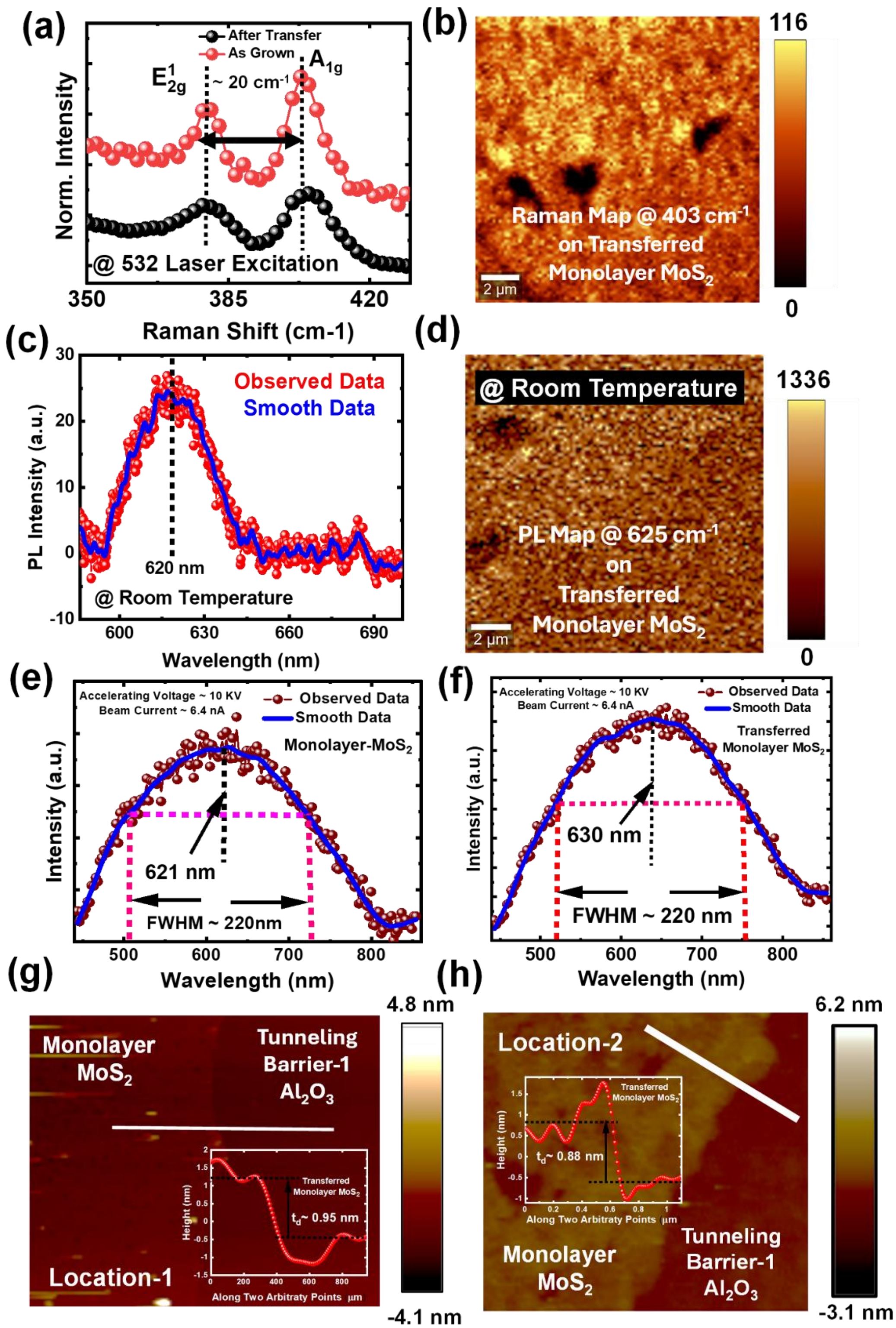

**Figure.1: (a) Raman spectroscopy for grown and transferred film, (b) Raman map at 403 cm$^{-1}$ for wet-transferred $MoS_2$ film, (c, d) Room temperature micro photo-luminescence characterization, (e, f) Cathodoluminescence characterization, (g, h) height estimation of the monolayer at two different locations**

Additionally, TMDC films are often partially oxidized, which further accelerates etching in B-HF. To mitigate these effects, a protective PMMA layer was spin-coated onto the $MoS_2$ film, followed by pre-baking at 110 °C for 180 s. The PMMA-coated sample was then immersed in the etchant solution. Within approximately 1–2 minutes, the PMMA/$MoS_2$ stack detached and floated on the solution surface. It was immediately transferred to a DI-water bath for rinsing and subsequently placed onto an $Al_2O_3$/Si substrate. Post-transfer baking was performed at 180 °C for 180 s, and the sample was allowed to dry under ambient conditions for 24 hours to enhance adhesion and ensure successful film transfer. Although this wet-transfer method enables transfer of large-area films, careful control of the B-HF to DI-water ratio and etching duration is critical to avoid film degradation. The transferred $MoS_2$ film was placed onto $Al_2O_3$/Si substrates, where $Al_2O_3$ served as the tunneling barrier material. A 10 nm thick $Al_2O_3$ layer was deposited by electron-beam evaporation at a deposition rate of ~0.02 Å $s^{-1}$, with an emission current of ~24 mA under high-vacuum conditions (~$1.19 \times 10^{-7}$ mbar). Atomic force microscopy (AFM) measurements were carried out at multiple locations to confirm the thickness and uniformity of the transferred $MoS_2$ layer (**Figure.1 (g, h)**)). The measured thickness ranged from 0.78 to 1.1 nm, consistent with monolayer $MoS_2$. A slight increase in surface roughness was observed after transfer, with the root-mean-square (RMS) roughness increasing from 0.429 nm to 0.533 nm. The surface morphology of the transferred film was further improved by (i) achieving lower initial roughness through controlled, slow CVD growth, (ii) post-growth annealing under vacuum or Ar/$H_2$ atmosphere, and (iii) optimizing the PMMA coating process using low spin speeds, longer spreading times, and gentle baking conditions (160–180 °C for 2–5 min). Film transfer to the target substrate remains a tedious process, as the $MoS_2$ layer can occasionally flip during transfer, leading to the appearance of multilayer regions (≥ 2 L) during material and device characterization. Such issues were carefully monitored and minimized in this work. To further validate that the grown film and the transferred film is monolayer, Raman characterization was carried out, where shift difference between two fundamental vibration modes ($E^1_{2g}$ and $A_{1g}$) comes out to be identical ~ 20 $cm^{-1}$ (as shown in **Figure.1 (a)**)[31–33]**.** To cross-check the uniformity of the transferred ML-film, Raman map was taken at peak position 403 $cm^{-1}$ for 400 $\mu m^2$ area (**Figure.1 (b)**). Indeed, minor pit holes were found, which can be avoided in lithographic process during device fabrication. Moreover, to observe any event of flipping of ML-film thereby possible exhibition of BL or Multi-layered characteristics, $A_{1g}$ peak i.e. 403 $cm^{-1}$ was chosen for Raman mapping since $A_{1g}$ mode gets more effected by layer numbers than $E^1_{2g}$ [34]. Similarly, photoluminescence (PL) was observed for as grown film with strong emission at peak location ~ 620 nm with FWHM ~ 32 nm, where mapping was done with equivalent significant counts at relatively shifted peak location ~ 625nm with proper uniformity, dominantly conceding with our claim for formation of few-S-vacancies in the monolayer as well as damage-free transfer of the same (see **Figure.2 (c, d)**)) [35,36] .To visit any other electronic excitations, cathode-luminescence (CL) characterizations at low accelerating voltages: 5-10 kV with beam current values: 1.6-6.4 nA were also carried out which again confirms that no-other defect state has been generated during transfer, however capping was used to avoid the burning of the film due to electron exposure. But it is worth mentioning that a minor shift (~ 9 nm, 28.4 meV in energy scale) in peak location was found between as grown and the transferred film as discussed in later section as shown in **Figure.1 (e, f)**). To estimate the elemental composition before and after transfer, X-ray Photoelectron spectroscopy (XPS) measurements were performed *i-situ* under ~ $10^{-9}$ mbar environment. For the direct CVD grown & transferred film, the positions of spin–orbit split peaks $2p_{3/2}$ and $2p_{1/2}$ (after proper carbon correction as shown in **Figure.2 (a, b)**) for S were found at 160.75 eV & 161.92 eV and 160 eV and 161.21 eV respectively, while in case of Mo, Mo-S peak locations are found to be at 227.83 ($3d_{5/2}$) & 231.062 eV ($3d_{3/2}$) and 227.80 ($3d_{5/2}$) & 231.061 eV ($3d_{3/2}$) respectively (see **Figure.2 (c-f)**))[37,38]. It is very interesting to note that Mo-O bonding was found for as grown film at 234.5 eV, which got completely vanished in transferred film due to O-removal by b-HF, thereby generating lots of vacancies along with the shift in doping characteristics with $\Delta E_V^F = 0.22eV \rightarrow 0.60eV$ (observed by valence band spectra, UPS method, **Figure.2 (g, h)**)). However, it is obligatory to mention that stoichiometry ratio of the as grown and transferred film were found to be 1:1.98 and 1:1.94 respectively. To fabricate isolated RTD devices (see top and cross-sectional schematic, **Figure.3 (a, d)**)), Laser writer was performed using S1813 positive photoresist with dose ~765 $\mu J/cm^2$ with 405 nm laser wavelength, to create Mesa structure of the transferred monolayer film. Opened locations were etched out via Reactive Ion etching (RIE) at 10W RF power using $SF_6$, $O_2$, $CHF_3$ and Ar with flow rates 8 sccm, 5 sccm, 5 sccm and 5 sccm respectively under 2 x $10^{-7}$ mbar vacuum pressure for 10 secs to avoid unwanted etching of first tunneling barrier [9,16]. To deposit the $2^{nd}$-TB, $Al_2O_3$ was deposited using e-beam at emission currents 24 mA at ~ 1.12 x $10^{-7}$ mbar vacuum environment as shown in **Figure.3 (b)**), while to create a window for deposition of bottom reservoir, RIE was further used with similar most configurations with etching time ~ 350 secs[9]. Unwanted F-contamination in Si was adequately taken care of. To finalize the device fabrication, Pt was deposited by e-beam via lithographic process. It is essential to mention that deposition of wafer scale continuous $Al_2O_2$ with least leakage condition was optimized before the fabrication of the final devices. However, it is essential to mention that the quality checks followed by the RTD fabrication were carried out for both atomic-layer deposited (ALD) and e-beam deposited $Al_2O_3$ thin tunneling barriers and found to be nearly equivalent within our desired applied voltage range. **Figure.3 (c)**) shows the band-diagram of the device near at K-valley of $MoS_2$, where electrons from n-Si resonantly inflow to $MoS_2$ QW due to energy quantization as discussed in the later sections.

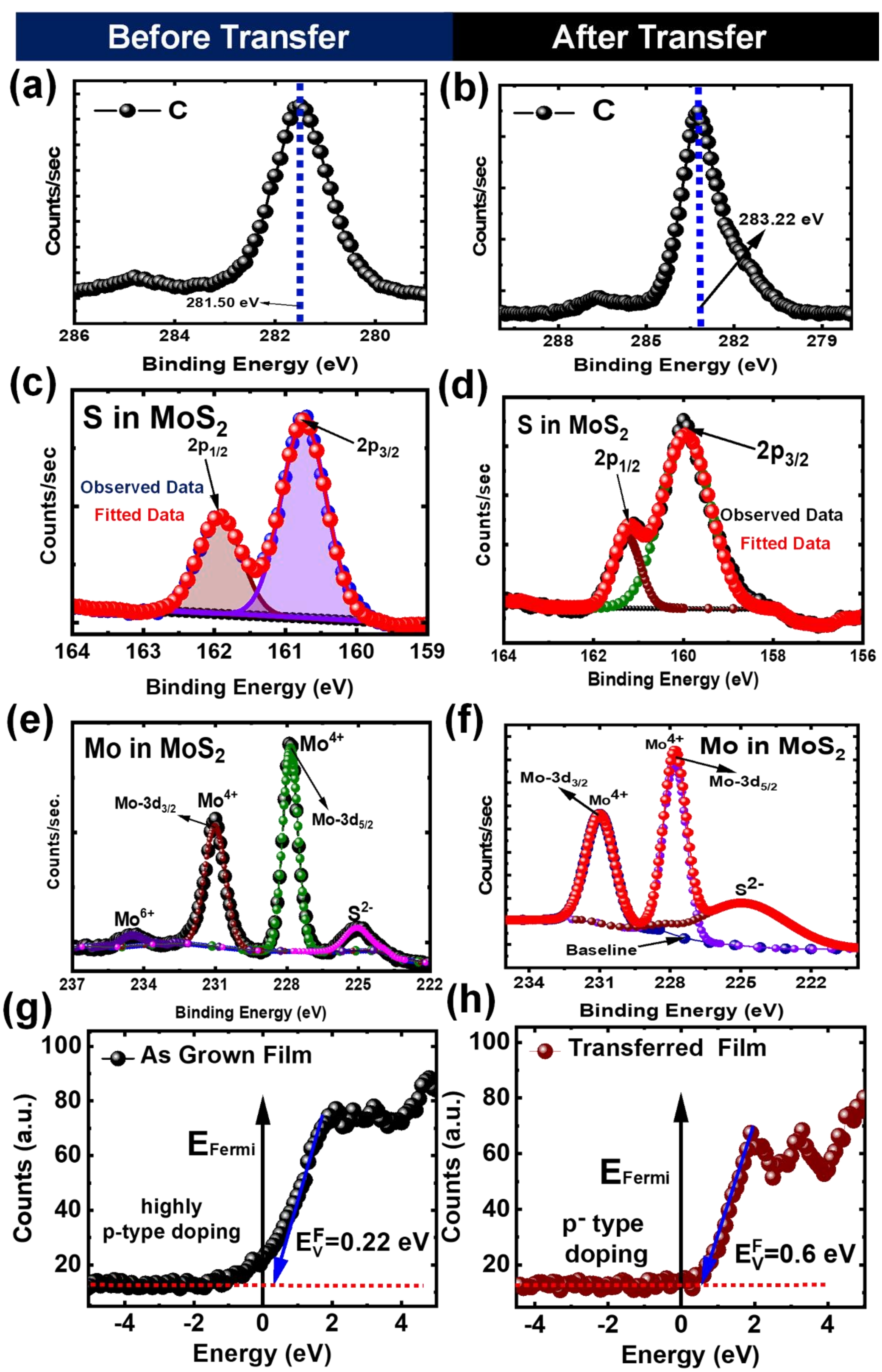

**Figure.2: High Resolution XPS spectra of the (a, b) Carbon, (c, d) S, (e, f) Mo and (g, h) valence band spectra for as grown and transferred films respectively.**

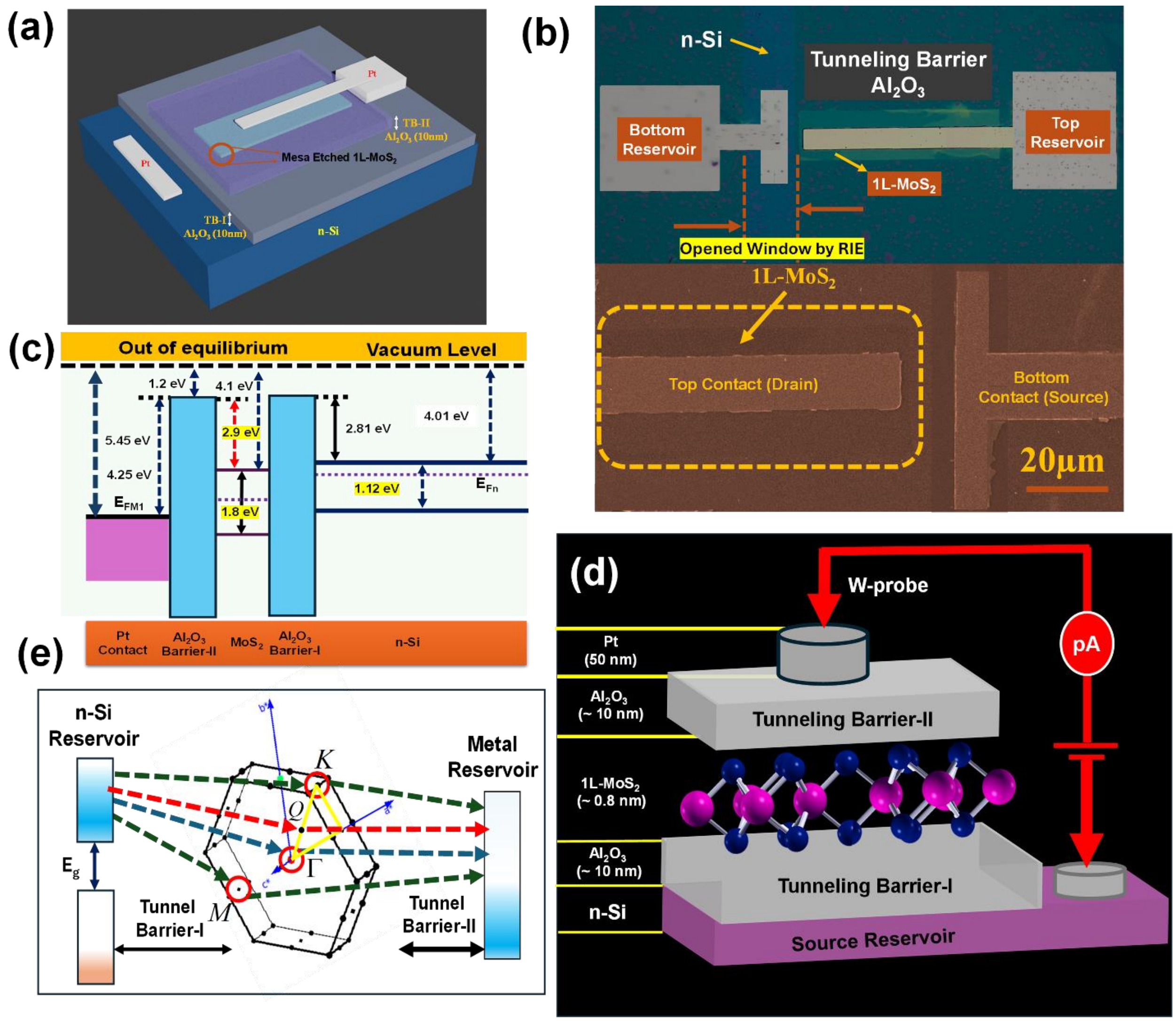

**Figure.3: (a) Top view of Device schematic, (b) Optical and FESEM Microscope images of the fabricated Devices, (c) Band diagram out of equilibrium condition near at $K$-point, (d) Cross-sectional view, (e) Different current components approaching from valley selective transport**

## III. Background Physics: Estimation of Band Structure & Transport Characteristics

To explore the fundamental insights in the transport of such monolayer based resonant tunneling devices with S-vacancies, a study based on Density Functional Theory (DFT) was carried out. Since, the CVD grown film was transferred via b-HF wet transfer method, formation of S-vacancies is quite obvious. To determine the equilibrium lattice constant, the in-plane lattice constant a = b was varied between 2.5 and 4.0 Å. The total energy was computed for each value, and the equilibrium lattice constant was identified as the value corresponding to the minimum energy in the energy curve.

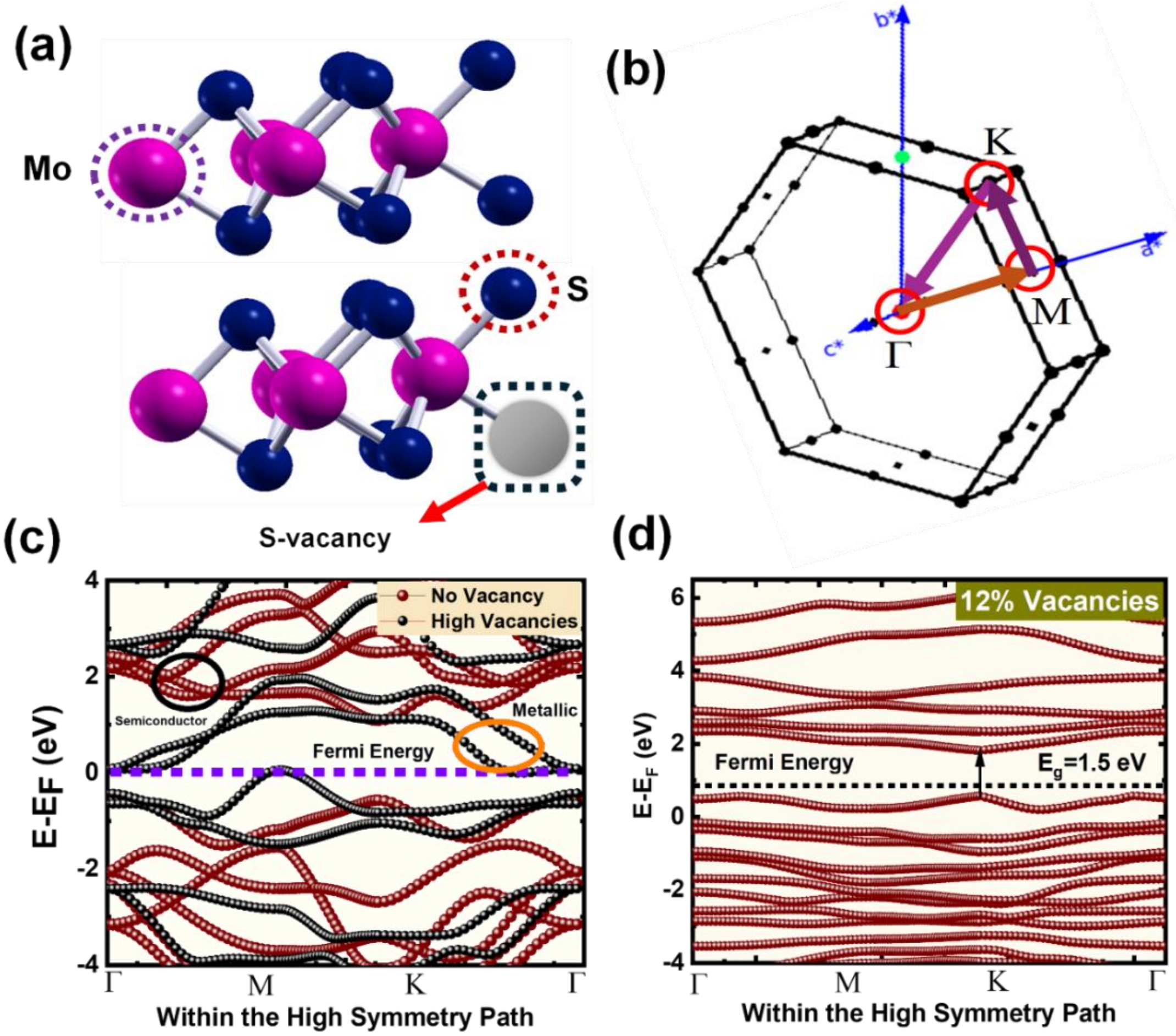

**Figure.4: (a) NxN Super-lattice designed for DFT Calculation with and without S-vacancy conditions,**

**(b) Considered High Symmetry Path in BZ, Band Structure (c) with no and 30% vacancies and (d) 12% vacancies**

For theoretical visualization of impact of S-vacancies in tranport properties, NxN $MoS_2$ super lattice with hexagonal symmetry (**Figure.4 (a)**) in monolayer configuration was first constructed with and without vacancies. To evaluate the band-structure, high symmetry path (HSP: $\Gamma \rightarrow M \rightarrow K \rightarrow \Gamma$, as shown in **Figure.4 (b)**) was chosen with appropiate cell parameters in free lattice format[39–41].

$$\vec{\tilde{A}}_1 = N(a,0,0)\ ;\ \vec{\tilde{A}}_2 = N\left(-a/2, \sqrt{3}a/2, 0\right)\ ;\ \vec{\tilde{A}}_3 = N(0,0,c)$$

Where, c corresponds to the vacuum gap to avoid spurious interactions and 'a' and 'N' being lattice constant and length of the cell respectively. From **Figure.4 (c),** it is clearly observable that $MoS_2$ monolayer lattice comes with direct-band gap (~ 1.9 eV to 1.8 eV) with 0 to 5 % vacancy condition. With the incerement of vacancies such as 12-16 %, band-gap decreases to ~ 1.6 to 1.5 eV. Once, vacancy exceeds 30 % or more (as mentioned as high vacancies), it loses its semiconductor nature and approaches towards metallic characteristics with band-gap appromiately 0.06 eV. It is noteworthy to mention that the modification in mobilty (hence in the effective mass) gets engrafted. Moreover, to evaluate the transport properties, band-gap values at various significant k-points within HSP were calculated with different S-vacancy conditions. For K-point grid sampling optimization, the grid density is a crucial parameter in DFT calculations, as it determines the resolution of reciprocal space sampling. Accurate integration over the Brillouin zone is essential for reliable total energy and electronic properties. A denser k-point grid improves resolution but increases computational time, necessitating a balance between accuracy and efficiency. To optimize the k-point grid, we systematically varied the grid density from 5 × 5 × 1 to 20 × 20 × 1 and calculated the total energy for each configuration. Convergence was achieved at a grid density of 10 × 10 × 1, where the change in total energy between successive grids was less than 0.001 eV. For the study of electronic properties, a denser grid is required to capture finer details, especially near high-symmetry points and critical regions in the band structure. Therefore, a k-point grid of 15 × 15 × 1 was selected for subsequent calculations. To elucidate the transport behavior, second quantized Hamiltonian was constructed considering the dominant interactions correlated with the monolayer film[5,9,28,42,43],

$$H = \sum_{i;k} H_{ISO}^{C;k} c_{i;k}^{+} c_{i;k} + \sum_{i,r;k,q,q'} \left\{ \tau_{ir}^{k,q,q'} c_{i;k}^{+} c_{r;q,q'} + \tau_{ri}^{k,q,q'} * c_{r;q,q'}^{+} c_{i;k} \right\}$$
$$+ \sum_{\alpha \in i,j;k,p} \left\{ \tilde{\tau}_{ij}^{\alpha;k,p} c_{i;k}^{+} c_{j;p} b_{\alpha,p} + \tilde{\tau}_{ji}^{\alpha;k,p} * c_{i;k}^{+} c_{j;k} b_{\alpha,p}^{+} \right\} \quad (1)$$

To formulate the device Green's function, the equations of motion of the system in second quantized forms are considered. The modified equations describing the system are given by[9,28],

$$\sum_{k} i\hbar \frac{d}{dt} c_{i;k} = \sum_{k} H_{k}^{ISO} c_{i;k} + \sum_{r;k,q} \tau_{ir}^{k,q} c_{r;q}$$
$$+ \sum_{\alpha \in j;k,p} \left\{ \tilde{\tau}_{ij}^{\alpha;k,p} c_{j;p} b_{\alpha,p} + \tilde{\tau}_{ji}^{\alpha;k,p} * c_{j;k} b_{\alpha,p}^{+} \right\} \quad (2.a)$$

$$\sum_{q} i\hbar \frac{d}{dt} c_{r;q}^{(2)} = \sum_{q} H_{R}^{ISO} c_{r;q}^{(2)} + \sum_{r;k,q} \tau_{ri}^{k,q*} c_{i;k}^{(2)} \quad (2.b)$$

$$\sum_{q'} i\hbar \frac{d}{dt} c_{r;q'}^{(1)} = \sum_{q'} H_{R}^{ISO} c_{r;q'}^{(1)} + \sum_{i;k,q'} \tau_{ri}^{k,q'*} c_{i;k}^{(1)}$$
$$+ \sum_{\alpha \in j;k,q'} \left\{ \tilde{\tau}_{rj'}^{\alpha;k,q'} c_{j';q'} b_{\alpha,q'} + \tilde{\tau}_{j'r}^{\alpha;k,q'} * c_{j';k} b_{\alpha,q'}^{+} \right\} \quad (2.c)$$

$$\sum_{k,k'} i\hbar \frac{d}{dt} b_{\alpha} = \sum_{k,k'} E_{\alpha}^{k,k'} b_{\alpha}^{k,k'} + \sum_{i,j} \left\{ \tilde{\tau}_{ij}^{\alpha;k,k'} * c_{i;k}^{+} c_{j;k'} + \tilde{\tau}_{ji}^{\alpha;k,k'} * c_{j;k'} c_{i;k}^{+} \right\} \quad (2.d)$$

(1, q) and (2, q') correspond to n-Si and metal reservoirs respectively. $\{i;k\}$ corresponds to canonically conjugate variables in determining the device Hamiltonian in both position (semi-continuous mode) and momentum space (discrete mode). The device Hamiltonian, $H_{D}^{ISO}$, in Eq. (1) can be expressed in effective mass approximation as,

$$H_D^{ISO}\left|\Theta\right\rangle = \sum_{k\in HSP}\left(\frac{1}{2}\sum \widehat{p}_i^k\left(m_k^{*-1}\right)_{ij}\widehat{p}_j^k + \phi(k;z)\right)\left|\Theta\right\rangle = \sum_{k\in HSP} E_m^k\left|\Theta\right\rangle \quad (3)$$

where, $\widehat{p}_i^k$ denotes the $i^{th}$ component of momentum; $\left(m_k^{*-1}\right)_{ij}$ denotes the $(ij)^{th}$ component of the inverse of effective mass tensor and $\phi(k;z)$ is the potential. Eq. (3) can be simplified as,

$$H\left|\Theta\right\rangle = \left(-\frac{1}{2m_{k;z}^*}\frac{\partial^2}{\partial z^2} - \frac{1}{2m_{k;y}^*}\frac{\partial^2}{\partial y^2} - \frac{1}{2m_{k;x}^*}\frac{\partial^2}{\partial x^2} + \phi(k;z)\right)\left|\Theta\right\rangle = E_m^k\left|\Theta\right\rangle \quad (4)$$

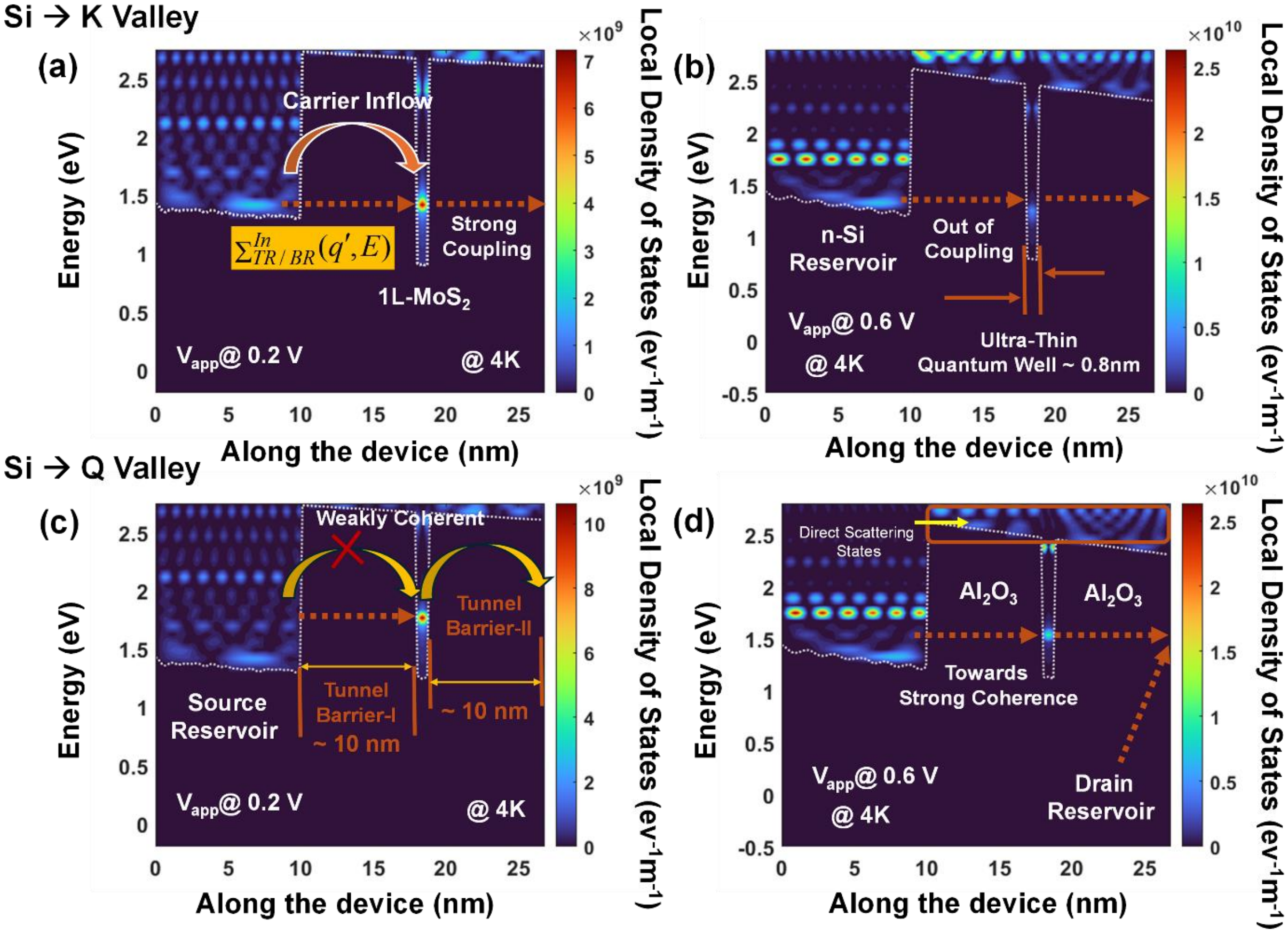

**Figure.5: LDOS contour plot depicting tunneling from Si through (a, b) K-valley and (c, d) Q-valley at applied bias 0.2 V and 0.6 V respectively**

From numerical perspective, the differentials in Eq. (4) can be written as finite difference equations. Consequently, Hamiltonian is a $(N_X N_Y N_Z)\times(N_X N_Y N_Z)$ matrix, where $N_X$, $N_Y$ and $N_z$ correspond to the number of grid points along each direction of discretization. To reduce the computational burden for such a Hamiltonian matrix, mode space approach is employed.

Here, the geometrical freedom of carriers perpendicular to 2D-$MoS_2$ sheet (x, y-directions) is treated separately from carrier transport (z-direction), leading to sub-bands, $\left|\zeta_n^k(x,y;z)\right\rangle$ with energies $E_n^k(z)$. Thus, the wavefunction for Eq. (4) can be written as[44],

$$\left|\Theta\right\rangle = \sum_{n,k} \xi_n^k(z) \left|\zeta_n^k(x,y;z)\right\rangle \tag{5}$$

Evaluated equations of motion for both momentum and position space for electrons have been solved in coupled mode space from the local non-equilibrium Green's function in the double barrier resonant tunneling medium coupled to various surroundings as given by[9,45,46],

$$\left[G^M(k,E)\right] = \begin{Bmatrix} \left[E^k I\right] - \left[H_{ISO}^k\right] - \left[\varepsilon_{k;l,y}^c + \varepsilon_{k;\lambda,x}^c\right]\delta_{l,l';\lambda,\lambda'}\delta(k-k') \\ -\left[\Sigma_M(q,E)\right] - \left[\Sigma_S(q',E)\right] - \left[\Sigma_{Scat}(k,k',E)\right] \end{Bmatrix}^{-1} \tag{6}$$

$\varepsilon_{k;l,y}^c$ and $\varepsilon_{k;\lambda,x}^c$ are the energy eigen values approaching from the lateral directions along which tunneling material offers degree of freedom.

**Table-I: Values of various parameters considered in quantum transport calculations[47–51]**

| Name of Parameters | Values |
|---|---|
| Electron affinity and Effective mass of $MoS_2$ | 4.1-4.2 eV and $0.46m_0$ |
| Di-electric constant of $MoS_2$ | ~ 2.6-2.9 |
| Phonon Deformation Potential for $MoS_2$ (Including Acoustic and Optical) | 3.5-8 eV |
| Lattice Constant of $MoS_2$ | 3.161 Å |
| $MoS_2$ Mass Density | 5.06 $g/cm^3$ |
| Electron affinity and Effective mass of Si | 4.01eV and 0.19 $m_0$ |
| Di-electric constant of Si | ~ 11.7 |
| Phonon Deformation Potential for Si (Including Acoustic and Optical) | 0.5-4 eV |

| Lattice Constant of Si | 5.43 Å |
| --- | --- |
| Si Mass Density | 2.33 g/cm$^3$ |
| $Al_2O_3$ electron affinity & effective mass | 1.1-1.2 eV and 0.35*$m_0$ |
| Di-electric constant of $Al_2O_3$ | 9.0 |

The local density of states (LDOS) occupied by electrons are therefore obtained to be[42],

$$[n(k,E)]=[G^n(k,E)]=\left(\frac{2}{2\pi a}\right)[G^M(k,E)]\left[\Sigma_M^{In}(q,E)+\Sigma_S^{In}(q',E)+\Sigma_{Scat}^{In}(k,k',E)\right][G^M(k,E)]^+ \quad (7)$$

Where,

$$\Sigma_{TR/BR}^{In}(q/q',E)=\left[\tau^{k,q/q'}\right]^+\left[n_{TR/BR}(k,E)\right]\left[\tau^{k,q/q'}\right]=\left[\tau^{k,q/q'}\right]^+\left[A_{TR/BR}(k,E)f_{TR/BR}(k,E,T)\right]\left[\tau^{k,q/q'}\right] \quad (8)$$

Notation $q/q'$ has been used to determine the metal and Si reservoir respectively with corresponding momentum value in the band-structure, from where electrons inflow/outflow to the active device. Different momentum dependent current components between different leads can be estimated to be[52,53]:

$$I_m(k,E)=\left(\frac{e}{2\pi\hbar}\right)Tr\left\{\overline{\overline{\Gamma}}_{ii'}^{S}(q',E)G_{i'j}^{M}(k,E)\overline{\overline{\Gamma}}_{jj'}^{M}(q,E)G_{j's}^{+M}(k,E)\right\} \quad (9)$$

Modification in self-energy matrices, thus modified broadening matrices are approaching from the recast of interaction between the Si-reservoir and the active device due to variation in the local band structure at the contact-domain as shown below[9].

$$\begin{aligned}\overline{\overline{\Gamma}}_{TR/BR}^{In}(E;q/q',l)&=\left(\overline{\overline{\Sigma}}_{TR/BR}^{In}(E;q/q',l)-\overline{\overline{\Sigma}}_{TR/BR}^{In}(E;q/q',l)^+\right)\\&=b_{q/q'}(k,q/q',l)\left(\Sigma_{TR/BR}^{i\rightarrow r}(E,q/q',l)-\Sigma_{TR/BR}^{i\leftarrow r}(E,q/q',l)^+\right)\end{aligned} \quad (10)$$

Here, *l* belongs the co-ordinates where junctions (e.g metal-insulator, insulator-semiconductor etc.) occurred. $b_{q/q'}$ corresponds to enhanced coupling factors being introduced to the active device. Hence, the measured resultant current comes out to be (see **Figure.7 (e)**),

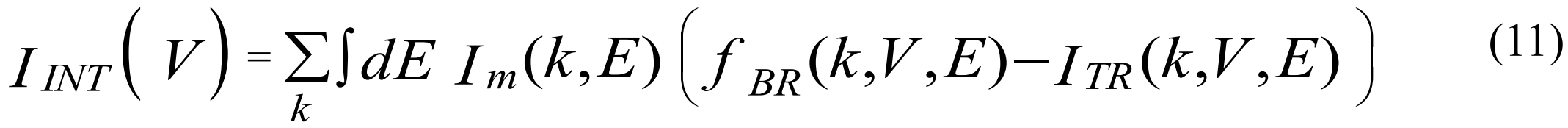

$$I_{INT}\left( V\right) = \sum_{k} \int dE \; I_m(k,E) \left( f_{BR}(k,V,E) - I_{TR}(k,V,E) \right) \qquad (11)$$

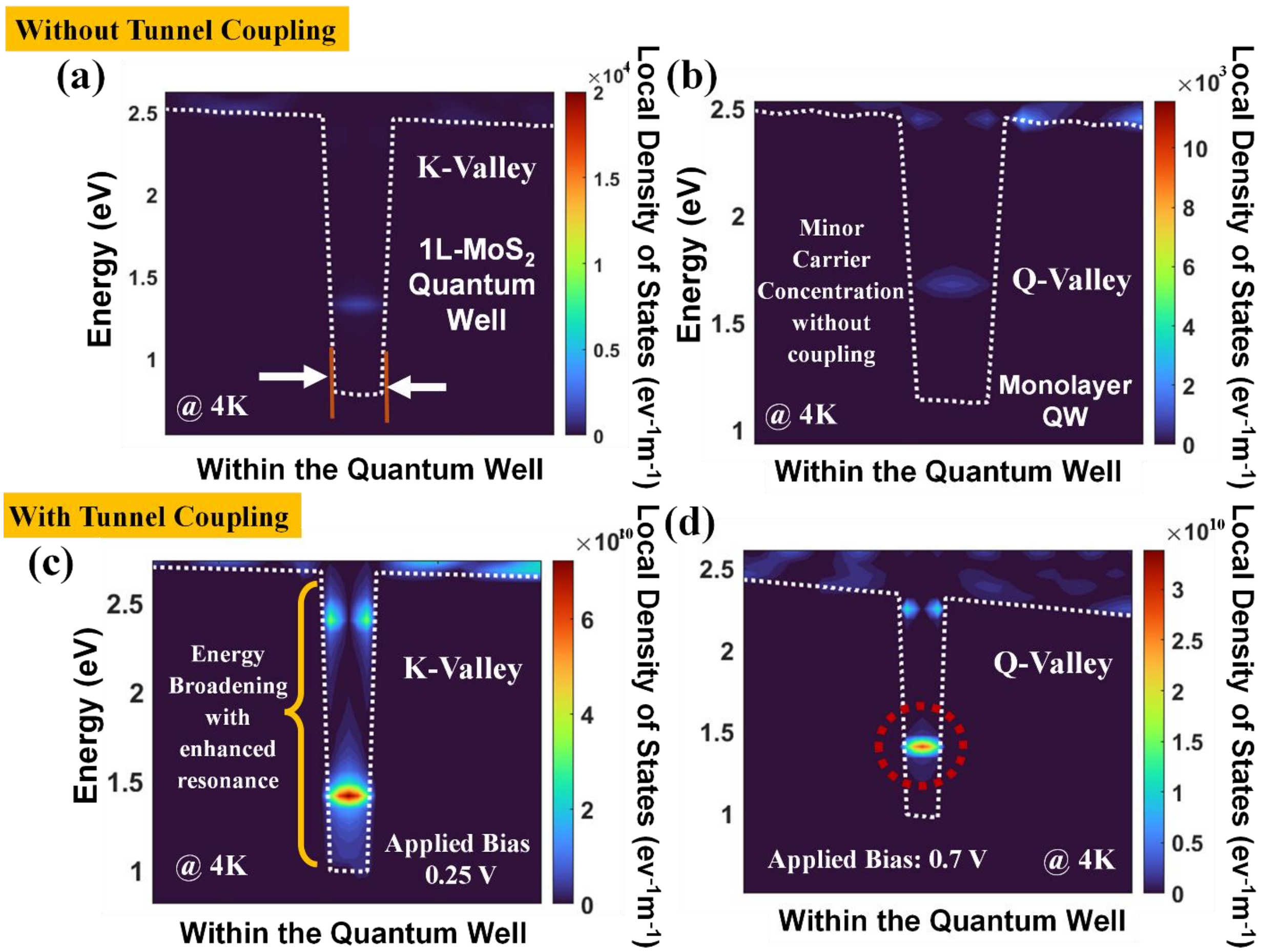

**Figure.6: (a, b) LDOS without coupling (c, d) TDOS with coupling at the 1L-$MoS_2$ Quantum Well with Si reservoir for K and Q valleys respectively.**

From the Local density of states (LDOS) plot (**Figure.5 (a-d)**), it is quite evident that at 0.2 V device reaches to 1st strong resonance through K-valley, gets out of phased at 0.6V, while via Q-valley 0.2 V remains weak coupling zone and reaches towards strong phase coherence above 0.6 V, near at 0.7 V at 4K environment. It is noteworthy to mention that even at 4K, the impact of phonon does not get diminished purely which can be visualized through LDOS plot where quasi quantized states in Si show intra-overlap. Such flexibility can be utilized as gate voltage tunable valley selective transport. To further elucidate the tunneling density of states (TDOS), it is important to note that the transferred film exhibits $p^-$ type doping, resulting in a high hole concentration and a significantly suppressed electron density of states (~$10^3$ $eV^{-1}$ $m^{-1}$), see **Figure.6**

**(a, b)**. However, the local density of states (LDOS) profile reveals that the electron density within the sandwiched monolayer $MoS_2$ rises sharply to ~$10^{10}$ $eV^{-1}$ $m^{-1}$ as depicted in **Figure.6 (c, d)**. This substantial enhancement is predominantly attributed to the inflow of tunneling carriers from the n-Si source reservoir, which effectively populates the quantized electronic states of different valleys in the $MoS_2$ layer despite its intrinsic p-type nature. Value of various parameters used in theoretical quantum transport calculations are listed in **Table-I**.

## IV. Electrical Measurement and Result & Discussion

The electrical characteristics of a single standalone double barrier RTD were measured using two tungsten (W) probe-micro tips attached to the Source Measure Units (SMU) inside the Lakeshore Cryotronic-336 probe station chamber under high vacuum at a pressure of ~ $10^{-6}$ mbars. The SMUs were connected with Keithley 4200A-SCS parameter analyzer[9]. To execute the measurement, W-micro probes were settled above the top and bottoms pad respectively, keeping a negative to positive voltage sweep over the top one. Initially in room temperature environment, a voltage sweep between -3 V to 3 V was given to experience the device response. Once the negative (differential) resistance (NDR) characteristics were found within range: -0.5 to 2.5 V, several runs were taken with very minute voltage-step ~ 0.5 mV, thus confirming the nature of NDR more prominently. To explore the variability of the NDR property from Cryogenic low temperatures to room temperature, Cryogenic Cyclic Refrigerator (CCR) was set towards 4K in the cryostat system. Once temperature reached 4K, I-V characterization was carried out thoroughly up to the room temperature as shown in **Figure. (7(a))**. It is noteworthy to mention that here multiple resonant tunneling peaks, approaching from different valleys were observed (**Figure.3(c-e))**. Notwithstanding, to cross check the device-to-device variability as well as to validate the retainability of valley selective resonant tunneling, the I-V characteristics for both Device-1 and 2 are plotted in **Figure. 7 (c, d)**. It is noteworthy to mention that at applied voltage > 2 V, carriers get started to flow above the tunneling barriers, thereby a huge thermionic emission current starts to dominate. Therefore, in Figure.6 (d), measurement was particularly carried out just to capture the tunneling feature and to avoid any voltage regarding damages. Similarity, in differential conductance plot (see **Figure.7(b)**), it is clearly observable that the first dominant resonant tunnel region (R-I) arrives from the transition: $X \rightarrow \mathrm{K}$, conveyed as nearly momentum conserved, where other two possible regions (R-II & R-III) are from $X \rightarrow Q$ and $X \rightarrow \Gamma$, depicted as weakly and non-momentum conserved respectively. To compare the performance of our 2D-RTD device with recently developed 2D and already matured Si, Ge, III-V based RTDs, resonant peak current & its density and peak to valley ration (PVR) have been measured.

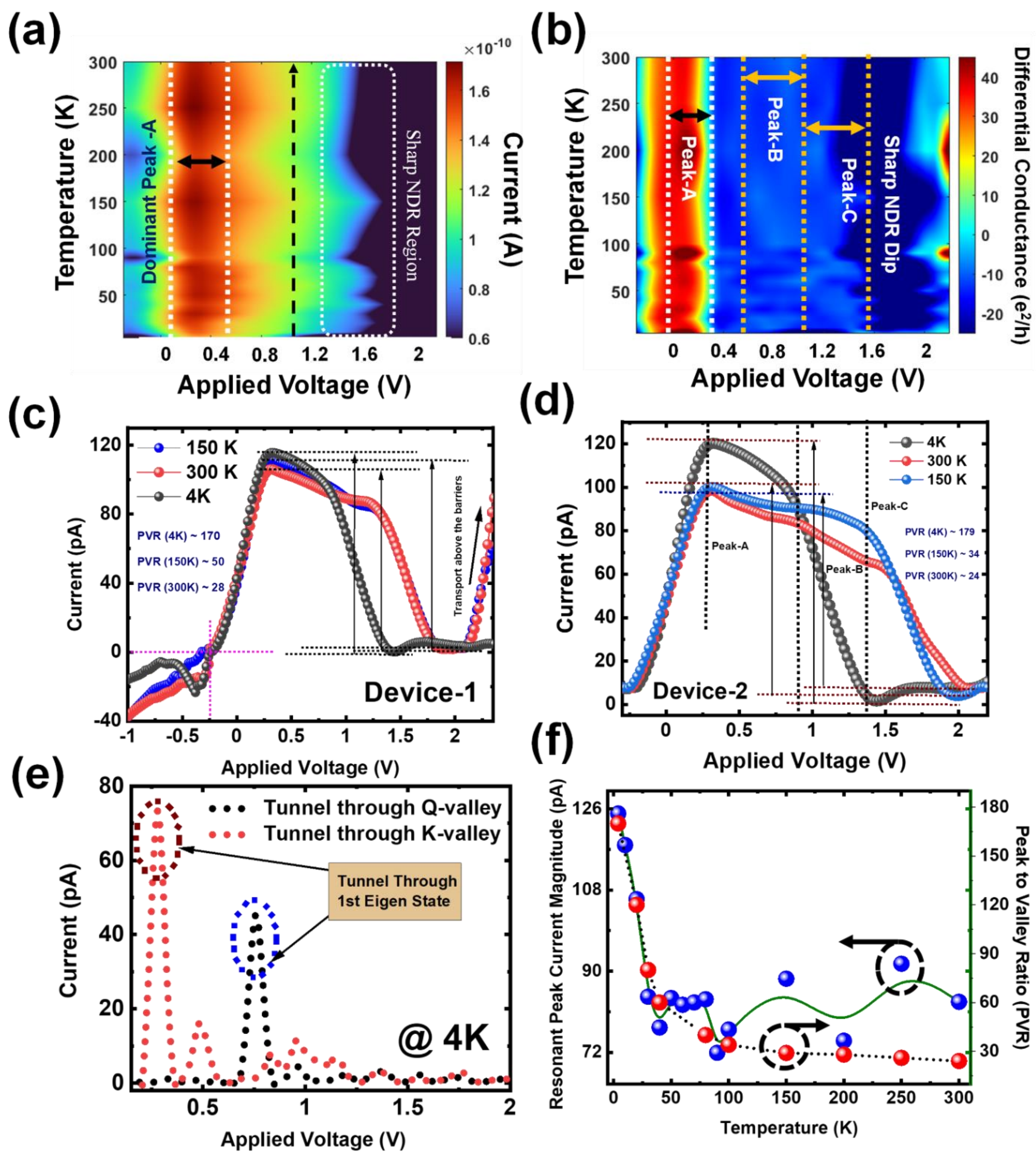

**Figure.7: Experimental (a) out of plane I-V characteristics 3D plot with various temperatures, (b) Differential conductance 3D Plot, (c, d) I-V plot to determine the PVR at 4 K, 150 K and 300 K for**

**Device-1 and 2 respectively , (e) Theoretically calculated conductance plot to justify valley selective transport with nearly 0.11 V FWHM, (f) Fluctuation of peak current magnitude (f) Peak to valley current ratio (PVR) vs temperature.**

Peak positions were also found to be oscillating in the voltage range: 0.28 to 0.32 V, imprinting the phononic dispersion replica in the transport contour plot, along the temperature. As illustrated in **Figure.7(f),** resonant peak current magnitudes were found to be ~ 124 pA (density, $J_d$ ~ 7.75 $\mu A/cm^2$) at 4K and 85 pA ($J_d$ ~ 5.25 $\mu A/cm^2$) at 300K with minor fluctuation at applied bias ~ 0.28 V purely captures the phonon's involvement.

| Tunneling Medium (Material) | Device Configuration | Peak to Valley Ratio (PVR) |
|---|---|---|
| III-V[54] | Double barrier architecture | Higher than 40 at room temperature |
| InP[55] | Tunnel diodes based on InP/InGaAs core-shell nanowires | ~ 10 |
| $WSe_2$[22] | Single barrier configuration (with hole tunneling) | 63.6 at 2 K and 16.2 at 300 K |
| Graphene[56] | Gate controlled Graphene-hBN-Graphene | 24 at 9K,<br>15 at 300K |
| 3L-$MoTe_2$[9] | Double barrier architecture (with electron tunneling) | 4 at 4K |
| InSe[10] | Gate controlled graphene/InSe/graphene structure (with electron tunneling) | Low |
| $WSe_2$[20] | Polarity-dependent twist-controlled resonant tunneling (with electron tunneling) | ~ 4.5 at 2K |
| 1L-$MoS_2$ **(This Work)** | Double barrier architecture (with electron tunneling) | 178 at 4K,<br>24 at 300K |

**Table-II: Comparison with recently developed 2D and other materials based RTDs**

Moreover, there was a significant response via demonstrating a huge PVR~ 178 at 4K and 24 at room temperature (**Figure.7(c)**), while theoretically it was expected to be ~ 200 at 4K (see **Figure.7 (e)**). Several devices were fabricated onto two different substrates to ensure the reported NDR characteristics. **Table II** demonstrates the progress made in the field of resonant tunneling devices (RTDs) during the past decade along the advancement done through this research work.

## V. Conclusion

In conclusion, we demonstrated the fabrication of 1L-$MoS_2$ based electron tunneling RTD architecture, via utilizing n-Si as the source reservoir and $Al_2O_3$ as the tunneling barriers. Comprehensive materials characterization verifies that the monolayer integrity is retained following wet transfer without any defects' contribution and subsequent fabrication processes, highlighting this methodology as a robust and scalable route for implementing ultra-thin 2D film–based quantum devices. For electron tunneling, we observed multiple resonant tunneling peaks approaching from various valley points (at $\mathrm{K}$, $Q$ and $\Gamma$) in $MoS_2$-monolayer lattice, with huge peak to valley current ratio (PVCR or PVR) ~ 179 at 4K and 24 at 300K. To explore the transport characteristics more fundamentally, non-equilibrium Green's function approach was extended to accommodate the momentum space contribution in the device performance, values of valley minimas were evaluated with and without S vacancies through Density Functional Theory. Such device may provide an efficient route as a coherent connector in between interacting qubits as well as in the development of Gate voltage tunable Spin-valley manipulation in Qubit via momentum conserved and non-conserved tunneling features. Notwithstanding, such device finds huge potential in the vast field of quantum oscillator, deterministic single photon emitter/detector as well as THz emitters.

## Acknowledgement

The authors acknowledge the Anusandhan National Research Foundation (ANRF), India, for supporting this work through the project *Development of Valley Qubit Technology Using Gated 2D Material Quantum Dot* (SPR/2023/000314). They also acknowledge the funding support from the National Quantum Mission (NQM), an initiative of the Department of Science and Technology (DST), Government of India. They extend their gratitude for the Ministry of Human Resource Development (MHRD), India, for funding this work through MoE-STARS/STARS-2/2023-061. The authors are grateful to Central Research Facility (CRF), Sophisticated Analytical & Technical Help Institutes (SATHI), Nanoscale Research Facility (NRF), IIT Delhi for infrastructural support. The authors thank Dr. Atul Kumar Singh, CRF, IIT Delhi and Dr. Manoj Kumar Kumawat, IIT Delhi, for having fruitful discussions.

## Authors' Contribution

#: AM and KS contributed equally to this work.

All authors contributed to data analysis and discussion of the results. All authors reviewed and contributed to the final version of the manuscript.

**Abir Mukherjee:** Conceptualization, Formal analysis, Investigation, Methodology, Software, Validation, Visualization, Writing – original draft. **Kajal Sharma:** Conceptualization, Methodology, Formal analysis, Visualization, Investigation, Writing–original draft. **Ajit K Katiyar**: Methodology, Writing – review & editing, **Saranya Das:** Formal analysis, Investigation**,** **Samit K Ray:** Resources, Funding acquisition, Supervision**,** **Samaresh Das**: Project administration, Investigation, Visualization, Resources, Funding acquisition, Supervision, Writing – review & editing.

## Data Availability Declaration

Data underlying the results and computer code to reproduce the theoretical results presented in this paper are not publicly available at this time but can be obtained from the corresponding author upon reasonable request.